\documentclass[reprint,superscriptaddress,amsmath,amssymb,aps,prb,twocolumn]{revtex4-1}

\usepackage{graphicx}
\usepackage{hyperref}

\begin{document}


\title{Threshold Interface Magnetization Required to Induce Magnetic Proximity Effect}

\author{O.\ Inyang}
\affiliation{Department of Physics, Durham University, South Road, Durham, DH1 3LE, United Kingdom}
\affiliation{Akwa Ibom State University, Mkpat Enin, Akwa Ibom State, Nigeria}

\author{L.\ Bouchenoire}
\affiliation{XMaS, The UK-CRG Beamline, ESRF, 71 Avenue des Martyrs, CS 40220, F-38043 Grenoble, France}
\affiliation{Department of Physics, University of Liverpool, Liverpool, L69 7ZE, United Kingdom}

\author{B.\ Nicholson}
\author{M.\ Toka\c{c}}
\altaffiliation[Present address:~]{Dept.\ of Fundamental Sciences, Alanya Alaaddin Keykubat University, Alanya-Antalya, Turkey}

\author{R.M.\ Rowan-Robinson}
\altaffiliation[Present address:~]{Dept.\ of Materials Science and Engineering, University of Sheffield, Sheffield, S1 3JD, United Kingdom}
\affiliation{Department of Physics, Durham University, South Road, Durham, DH1 3LE, United Kingdom}
\author{C.J.\ Kinane}
\affiliation{ISIS Neutron Facility, STFC Rutherford Appleton Laboratory, Didcot, OX11 0QX, United Kingdom}
\author{A.T.\ Hindmarch}
\email[Email:~]{a.t.hindmarch@durham.ac.uk}
\affiliation{Department of Physics, Durham University, South Road, Durham, DH1 3LE, United Kingdom}

\date{\today}


\begin{abstract}
Proximity-induced magnetization (PIM) has broad implications across interface-driven spintronics applications employing spin-currents. We directly determine the scaling between PIM in Pt and the temperature-dependent interface magnetization in an adjacent ferromagnet (FM) using depth-resolved magnetometry. 
The magnetization due to PIM does not follow the generally expected linear scaling with the FM interface magnetization, as a function of temperature. 
Instead, it vanishes whilst the FM interface magnetization remains. %
The effective magnetic susceptibilities of heavy metal (HM) layers are shown to give rise to the previously unexplained asymmetric PIM found in HM/FM/HM trilayers.

\end{abstract}
\maketitle

\section{Introduction}

Platinum, and other heavy metals (HM), are widely used for interface-driven spintronics applications. 
Proximity-induced magnetization (PIM) arises when these materials are placed in direct contact with a ferromagnetic (FM) material. %
PIM has been implicated in many spintronic phenomena including the spin-Hall effect~\cite{Zhang2015}\ 
and spin-Hall magnetoresistance~\cite{Huang2012,
Lu2013}, 
the spin-Seebeck~\cite{Kuschel2015}\ 
and anomalous-Nernst~\cite{Guo2014}\ effects, 
the anomalous-Hall effect~\cite{Shimizu2013}, 
spin-relaxation~\cite{Freeman2018},
interface spin-transparency~\cite{Zhang2015a,Avci2015}\ 
and spin-pumping~\cite{Caminale2016}, 
interfacial Dzyaloshinskii-Moriya interaction~\cite{Rowan-Robinson2017}\ and chiral domain wall motion~\cite{Ryu2014},
and magnetization ($M$) reversal driven by spin-orbit torques~\cite{Peterson2018}\  
and electric fields~\cite{Miwa2017}. 

The first report of PIM in Pt suggested that it is due to interfacial hybridization between the Pt $5d$ band and the $3d$ band of FM Fe, Co, Ni, and their alloys~\cite{Schutz1990}. This remains the basic physical description of the PIM mechanism~\cite{Nakajima1998,Liang2016,Rowan-Robinson2017}. 
Element specific magnetometry using x-ray magnetic circular dichroism and x-ray resonant magnetic reflectivity (XRMR) have shown that PIM typically decays away from the interface over a lengthscale of $\sim 1$~nm~\cite{wilhelm2000layer,Geissler2001}, and in some cases suggest surprisingly large amplitudes up to $\sim 0.6 \text{ }\mu_\mathrm{B}/$atom 
in nominally non-magnetic materials~\cite{Antel1999,Kuschel2015,Klewe2016}. %
Pt on YIG, a ferrimagnetic oxide, may~\cite{Lu2013}\ or may not~\cite{geprags2012investigation}\ support PIM. Despite the breadth of topics where magnetic proximity effects are implicated, little detail is known about the factors determining PIM at a HM/FM interface.

Here we demonstrate how the FM interface magnetization modifies PIM. We use a designed model structure consisting of a FM alloy with tailored Curie temperature, sandwiched between Pt layers. Complementary depth- and element-resolved magnetometries enable simultaneous extraction of the relationship between PIM and the temperature-dependent FM magnetization at both interfaces. This approach provides new insights into the phenomenology of PIM.

Pt is a paramagnet which is close to the Stoner instability, $Ug(E_\mathrm{F}) = 1$, where $U$ is the electron-electron Coulomb (exchange) interaction and $g(E_\mathrm{F})$ the electronic density of states at the Fermi energy. Such `Stoner-enhanced' paramagnets have magnetic susceptibility
\[\chi = \frac{M}{H} = \frac{\chi_\mathrm{P}}{1-Ug(E_\mathrm{F})} = \chi_\mathrm{P}S,\]
where $H$ is the applied magnetic field and $\chi_\mathrm{P} = 2\mu_0\mu_\mathrm{B}^2g(E_\mathrm{F})$ is the (approximately magnetic field- and temperature- independent) Pauli susceptibility, with $\mu_0$ the vacuum permeability and $\mu_\mathrm{B}$ the Bohr magneton. In Pt the Stoner enhancement factor $S = 1/[1-Ug(E_\mathrm{F})]$ typically takes a value $S \approx 4$~\cite{Herrmannsdoerfer1996}.

The conventional description of PIM induced at the interface between a HM and a FM can be encapsulated within a simple phenomenological model~\cite{Lim2013, hase2014proximity},
\begin{equation}\label{MvM}
\mathrm{PIM} = \chi H_\mathrm{ex}
= \chi_\mathrm{P}SJ M_\mathrm{FM},
\end{equation}
whereby, e.g., through 3$d$-5$d$ hybridization the interface magnetization, $M_\mathrm{FM}$, in the FM and (dimensionless) interfacial exchange-coupling, $J$, produce a local `exchange field', $H_\mathrm{ex}$~\cite{Lim2013}, that results in a spontaneous equilibrium PIM in the HM. Roughness and intermixing modify $J$, intermixing of the HM into the FM modifies $M_\mathrm{FM}$, and intermixing of the FM into the HM modifies $\chi$. The microscopic physical structure of the interface is therefore important in determining PIM~\cite{Ferrer1997,Tokac2015,Liang2016}.

Studies of the scaling between PIM and the FM interface magnetization are limited. Indirect PIM measurements on Pd/Fe/Pd multilayers (Pd is also a Stoner-enhanced paramagnet), suggest PIM vanishes despite Fe remaining strongly magnetized~\cite{Cheng2004}. Direct measurements of PIM for ultra-thin Fe sandwiched between Pd show PIM following the temperature dependence of a 3D FM rather than the quasi-2D behaviour of the Fe layer~\cite{hase2014proximity}. In thicker Pt/Co/Pt~\cite{Rowan-Robinson2017}\ and Pd/Co/Pd~\cite{Kim2016}\ trilayers the PIM at the two interfaces differs: PIM is larger at the upper FM/HM interface than at the lower HM/FM interface. In these cases Eq.~\ref{MvM}\ does not adequately describe the observed behaviour.

\section{Experimental Approach}

Here we describe an approach to disentangle the PIM at an interface between a FM and HM using a combination of neutron and x-ray reflectivity measurements.

Polarised neutron reflectivity (PNR) is sensitive both to the physical structure of the sample via the nuclear scattering of neutrons, and enables quantitative depth-resolved extraction of the local magnetization via the interaction between the neutron spin and the local magnetic induction~\cite{Ankner1999}. As the PIM in the HM consists of small magnetic moments that are spatially concentrated close to the interface with the FM, the magnetic depth profile measured by PNR is sensitive primarily to magnetism in FM.

One the other hand, to unambiguously extract the small, spatially concentrated PIM in HM, we use x-ray resonant magnetic reflectivity (XRMR). Tuning the energy of the incident circularly-polarised x-ray beam close to an appropriate elemental absorption edge, in this case Pt $L_3$, provides depth-resolved element selective magnetic sensitivity --- as with other x-ray techniques XRMR is sensitive to the overall physical structure via the local electron density through Thompson scattering, but most importantly, the resonant scattering interaction here is sensitive to magnetism only in Pt~\cite{Macke2014}.

The combination of these approaches enables 
the (large) interface magnetization in FM at FM/HM interfaces to be determined quantitatively from PNR, and 
the (much smaller) PIM in the HM on the other side of the same interface to be determined with element specificity from XRMR. The HM magnetization (PIM) is very much smaller than the FM magnetization; so the magnetization in HM is essentially invisible to PNR while the magnetization in FM is entirely invisible to XRMR. The depth selectivity of these scattering techniques further enables us to investigate PIM and interface magnetization at multiple interfaces in a multilayered sample within a single set of measurements. By studying both interfaces as a function of temperature, we aim to determine the amplitude of the PIM on the Pt side of the interface as the FM magnetization on the other side of the same interface varies over a wide range, in a single sample.

Here, a Pt(3~nm)/Co$_{28}$Fe$_{28}$Ta$_{30}$B$_{14}$ (10~nm)/Pt (3~nm) trilayer film was used to clarify the linkage between PIM and FM interface magnetization. The subscripts represent the nominal composition of CoFeTaB (CFTB), and 30 at.~\%\ Ta reduces the Curie temperature, $T_\mathrm{C}$, to below room temperature. Experimental details can be found in the supplementary material~\cite{suppl}. We have found that Ta can diffuse vertically within this amorphous alloy during film deposition, and that subtle vertical diffusion of Ta within CFTB causes variation in the local $M$ and $T_\mathrm{C}$~\cite{tokacc2017temperature}. This is similar to Fe$_{(100-x)}$Ta$_x$ amorphous alloys, where $T_\mathrm{C}$ can be tuned from $\sim 200$~K to below 100~K by increasing $x$ from $\sim 15$~\%\ to $\sim 30$~\%~\cite{Fukamichi1981}. 
Here Ta diffusion produces regions of different composition through the thickness of the CFTB film where the upper and lower interfaces have different $T_\mathrm{C}$, further enabling independent study of PIM at each interface.

\section{Results and Discussion}

\begin{figure}
\centering
\includegraphics[width=0.9\linewidth]{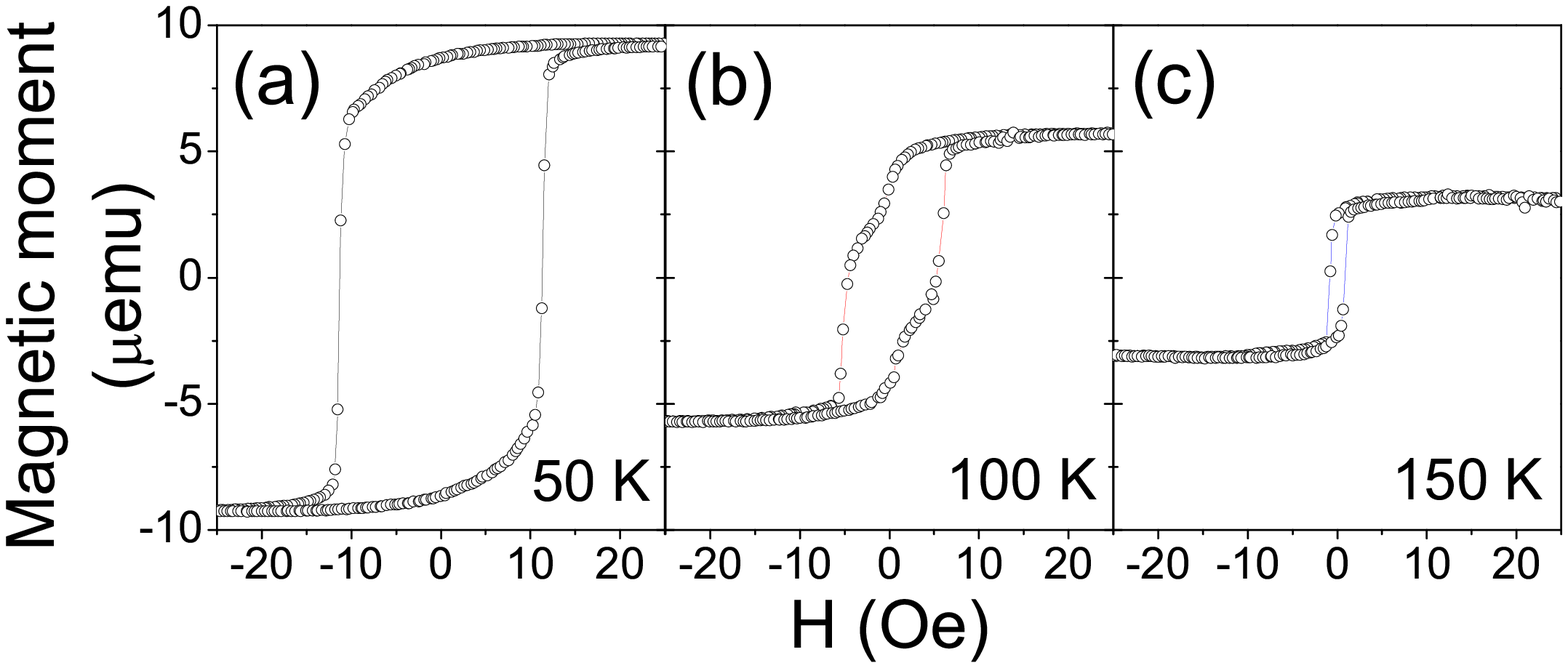}
\vspace{2mm}\\
\includegraphics[width=0.9\linewidth]{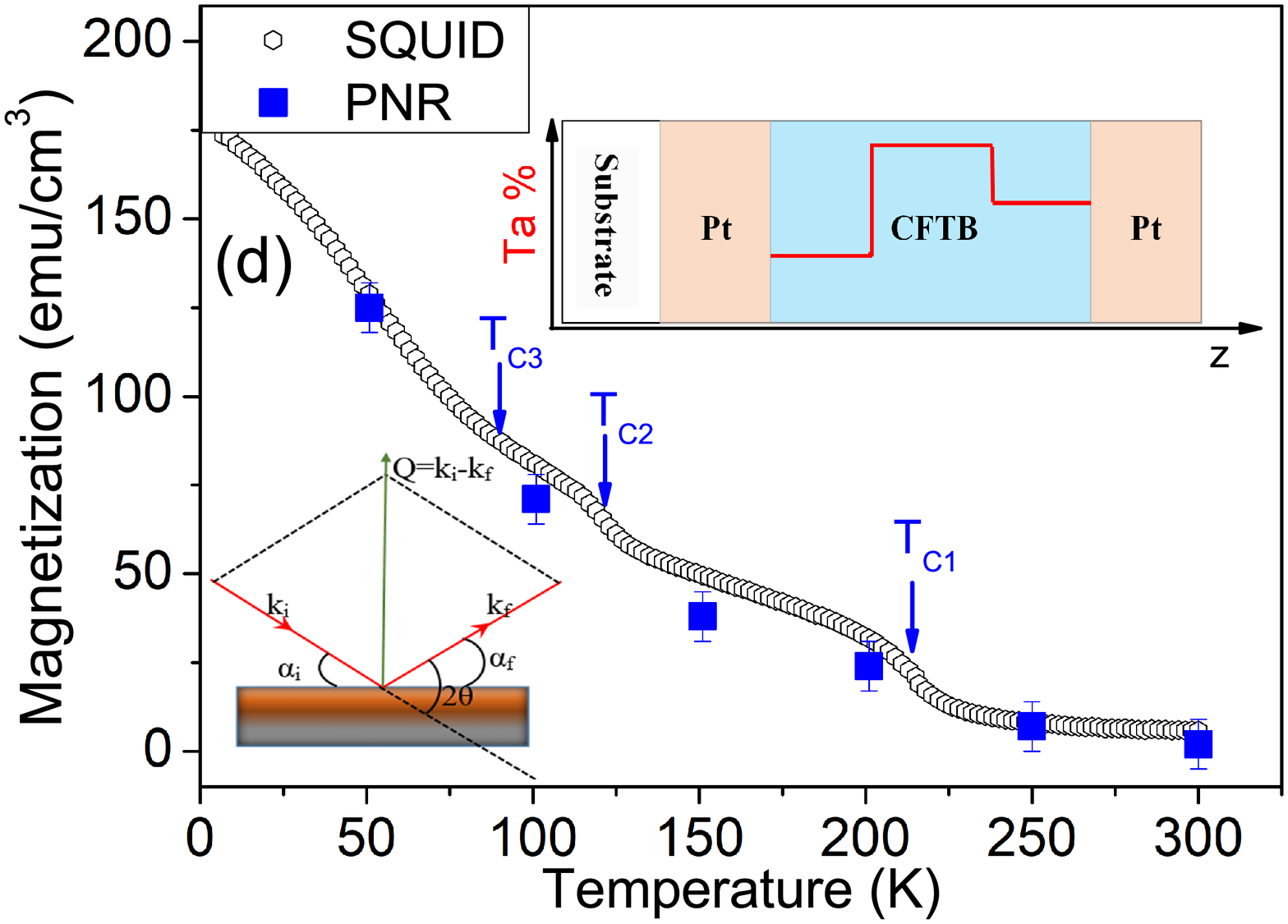}
\caption{Temperature dependent magnetization (main) for trilayer Pt/CFTB/Pt (open markers), showing three transition temperatures $T_\mathrm{C1-3}$ indicated with arrows. Filled markers are extracted from PNR measurements. Magnetization hysteresis loops show two-step magnetization reversal at temperatures between $T_\mathrm{C3}$ and $T_\mathrm{C2}$ (b), and single-step reversal both below $T_\mathrm{C3}$ (a) and between $T_\mathrm{C2}$ and $T_\mathrm{C1}$(c). Inset schematics show (left) the scattering geometry for PNR and XRMR measurements, and (right) the sample layer structure and tantalum distribution profile through the CFTB layer.}\label{fig:SQUID3}
\end{figure}

The in-plane saturation volume $M(T)$, measured by superconducting quantum interference device (SQUID) magnetometry
after field-cooling 
is shown in the main frame of Fig.~\ref{fig:SQUID3}. The Pt magnetization, $M_\mathrm{Pt}$, due to PIM is expected to be small and localized at the interfaces, so the volume magnetization is expected to be dominated by $M_\mathrm{CFTB}$, the magnetization in CFTB. As expected, $M$ decreases with increasing $T$. The data show that the CFTB layer does not act as a single homogeneous magnetic slab with a single $T_\mathrm{C}$. Instead it suggests three transitions denoted as $T_\mathrm{C1}\sim 215$~K, $T_\mathrm{C2}\sim 120$~K and $T_\mathrm{C3}\sim 80$~K in Fig.~\ref{fig:SQUID3}. These 
transitions 
confirm the anticipated compositional variation through the CFTB layer thickness.


This compositional variation is further confirmed by magnetic hysteresis loop measurements, made using SQUID vibrating sample magnetometry (SQUID-VSM), with examples shown in Fig.~\ref{fig:SQUID3}(a), (b) and (c). At $T < T_\mathrm{C3}$ a single-step reversal was observed, shown at $T=50$~K in Fig.~\ref{fig:SQUID3}(a). When $T_\mathrm{C3} < T < T_\mathrm{C2}$ a distinct two-step reversal of $M$ was observed, shown at $T=100$~K in Fig.~\ref{fig:SQUID3}(b). This indicates two magnetic regions within the CFTB layer, with different coercivities. As $T$ increased the signal amplitude from the region with smaller $H_\mathrm{C}$ decreased (see supplementary material, Ref.~\onlinecite{suppl}), vanishing at $T_\mathrm{C2}\sim 120$~K. At $T > T_\mathrm{C2}$ a single-step reversal was observed, shown at $T=150$~K in Fig.~\ref{fig:SQUID3}(c).

Assuming the CFTB film comprises three sub-layers each with slightly different Ta concentrations, the two-step reversal of $M$ observed for $T_\mathrm{C3} < T < T_\mathrm{C2}$ indicates that $T_\mathrm{C3}$ corresponds to the $T_\mathrm{C}$ of the sub-layer in the center of the film. At $T<T_\mathrm{C3}$ the film acts as a single magnetic entity, but at $T_\mathrm{C3} < T < T_\mathrm{C2}$ the upper and lower FM CFTB sub-layers are magnetically decoupled by the paramagnetic central CFTB sub-layer, similar to Ref.~\onlinecite{magnus2016long}. The Ta concentration profile producing this behaviour is shown schematically inset to the upper right of Fig.~\ref{fig:SQUID3}(d) by the solid line, where we presuppose that $T_\mathrm{C2}$ and $T_\mathrm{C1}$ correspond to the $T_\mathrm{C}$ of the upper and lower CFTB sub-layers, respectively.

\begin{figure*}
\centering
\includegraphics[width=0.88\linewidth]{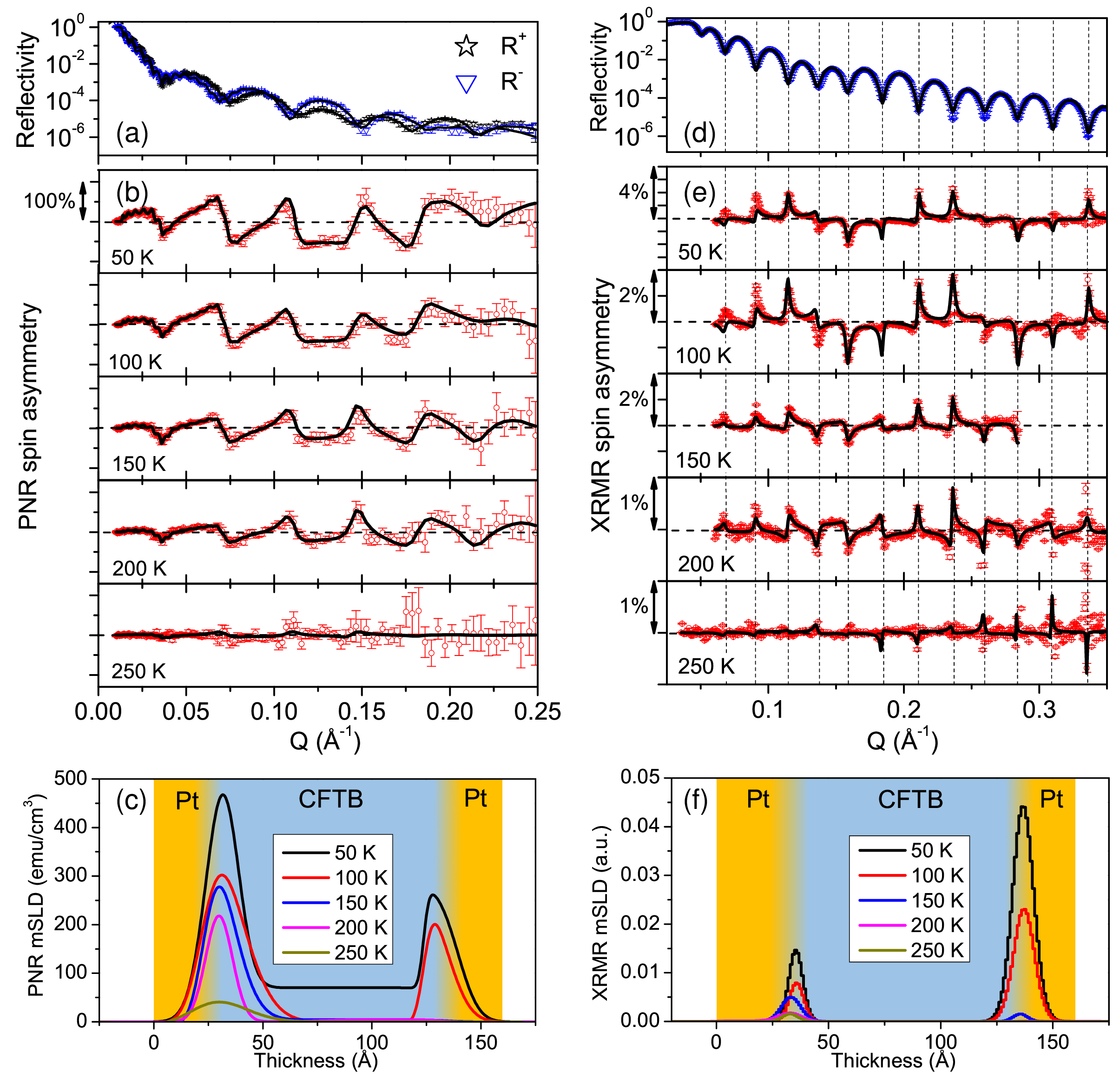}
\caption{Example reflectivity, and spin-asymmetry at various temperatures, for 
Pt/CFTB/Pt from (left) PNR and (right) XRMR. The horizontal dashed lines show zero spin-asymmetry, and vertical dashed lines in (d) and (e) highlight that the peaks in the XRMR spin-asymmetry are strongest at Kiessig fringe minima in the reflectivity. The solid lines show best-fits to the reflectivity and spin-asymmetry, corresponding to the mSLD profiles in (c) and (f). The sample layer structure is indicated against the mSLD profiles; slight structural differences, primarily in the lower Pt layer thickness, may result from the different sampling volumes (beam spot sizes) for PNR ($\sim$~3~cm~$\times$~3~cm) and XRMR ($\sim$~200~$\mu$m~$\times$~300~$\mu$m), however, layer thicknesses extracted from unconstrained fitting of PNR and XRMR data are consistent within standard error. Apparent overlap of the PNR mSLD profile into Pt layers, and XRMR mSLD profile into CFTB, is a result of the error-function profile used to model the interface roughnesses.
}\label{fig:ReflandSA}
\end{figure*}

\begin{figure}
\centering
\includegraphics[width=0.85\linewidth]{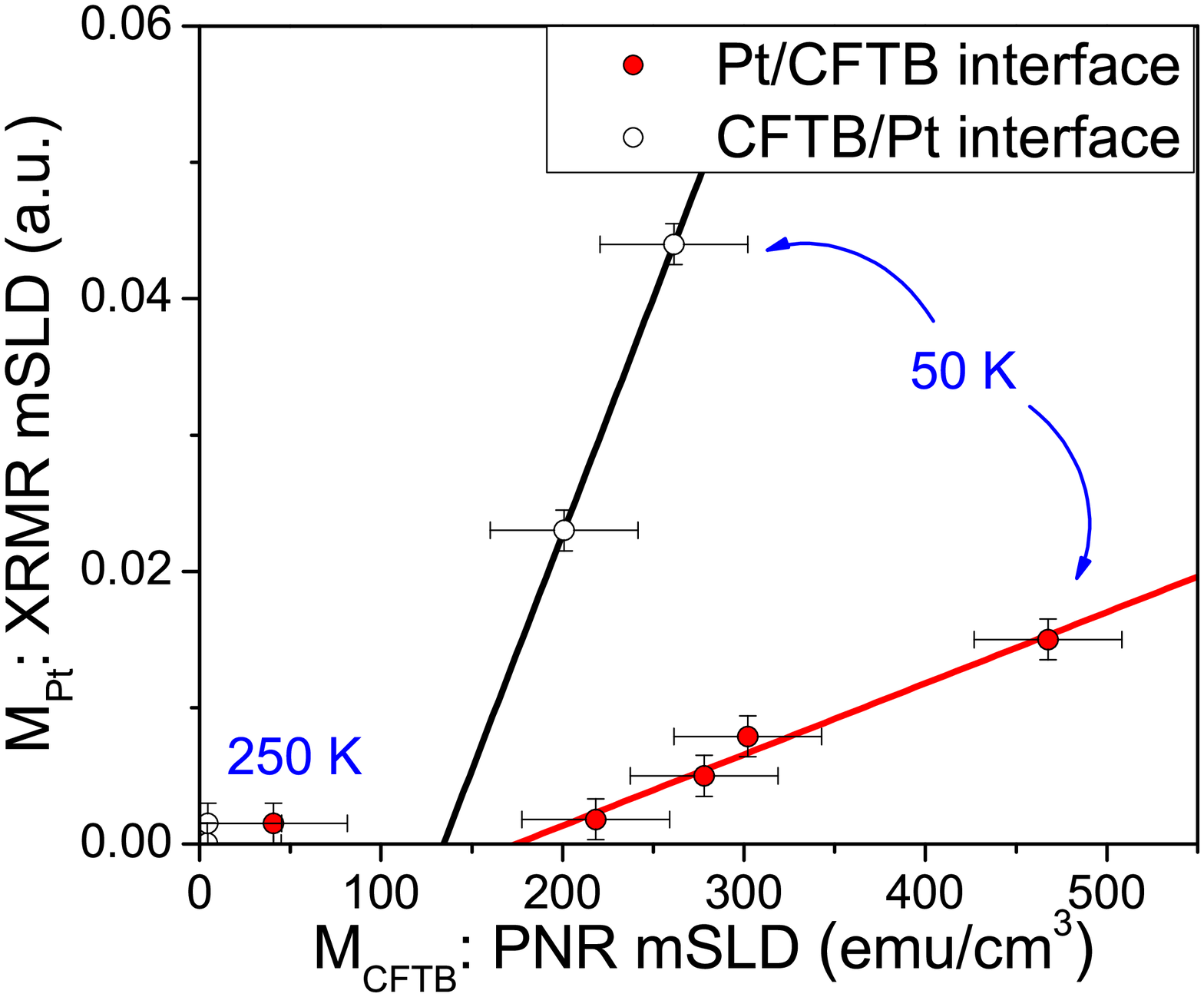}
\caption{Parametric scaling of PIM in Pt (XRMR mSLD) against CFTB interface magnetization (PNR mSLD) for lower Pt/CFTB (filled markers) and upper CFTB/Pt interfaces as a function of temperature. Solid lines show the linear scaling between PIM and interface magnetization over the temperature range where each interface is locally ferromagnetic. The conventional linear description of PIM, Eq.~\ref{MvM}, necessarily passes through the origin.}\label{fig:PIMvsM}
\end{figure}

A quantitative description of the depth dependence of $M_\mathrm{CFTB}$ as a function of $T$ was obtained using polarised neutron reflectometry (PNR). The approach used to extract the magnetic scattering length density (mSLD) profile is described in Ref.~\onlinecite{suppl}. %
PNR is found here to be sensitive primarily to magnetism in CFTB due to the small PIM magnetic moment in Pt. Fig.~\ref{fig:ReflandSA}(a) shows PNR at $T=50$~K. The PNR spin-asymmetry in Fig.~\ref{fig:ReflandSA}(b) shows the normalized difference between the reflectivity channels $R^+$ and $R^-$ as a function of scattering vector, $Q$, for different $T$. The scattering geometry is schematically illustrated in the lower left inset of Fig.~\ref{fig:SQUID3}(d). $R^+$, $R^-$, and the derived spin-asymmetry at each $T$ were fitted simultaneously using \texttt{GenX}\ software~\cite{bjorck2007genx}, each using the same structural scattering length density (sSLD) model for the sample derived from x-ray reflectivity (XRR). Best-fits, determined from reduced-$\chi^2$, are shown as solid lines in Fig.~\ref{fig:ReflandSA}(a) and (b). The CFTB layer was modelled as three sub-layers, as suggested by SQUID(-VSM) measurements shown in Fig.~\ref{fig:SQUID3}. This was required to describe the PNR spin-asymmetry (see Ref.~\onlinecite{suppl}), and adding further sub-layers provided no significant improvement to the fits. All reduced-$\chi^2$ values are below 1.4. 

The mSLD profiles obtained from best-fits to the PNR data are shown in Fig.~\ref{fig:ReflandSA}(c). The mSLD profile at $T=50$~K shows a uniform $M$ in the center of the CFTB layer, with larger $M$ near the upper interface, and still larger $M$ near the lower interface. At $T=100$~K the central region is paramagnetic, confirming our interpretation that $T_\mathrm{C3}$ corresponds to the $T_\mathrm{C}$ of the central CFTB sub-layer, with slightly higher Ta concentration. At $T=150$~K the upper CFTB sub-layer becomes paramagnetic, demonstrating that $T_\mathrm{C2}$ indeed corresponds to the $T_\mathrm{C}$ of the upper sub-layer. The mSLD of the lower CFTB sub-layer decreases with increasing $T$. The best-fit at $T=250$~K, i.e., above $T_\mathrm{C1}$ where the entire CFTB film is paramagnetic, suggests a small mSLD in the lower CFTB sub-layer. This is indicative of the uncertainty in $M_\mathrm{CFTB}$ extracted from PNR. Integrating each mSLD profile gives the volume $M_\mathrm{CFTB}$, plotted as filled square markers in Fig.~\ref{fig:SQUID3}(c); the volume $M_\mathrm{CFTB}$ obtained from unconstrained fitting of PNR is consistent with the measured $M(T)$ within experimental uncertainty.

The depth dependence of PIM in the Pt layers was extracted from XRMR measurements at the Pt L$_3$ absorption edge as a function of $T$. Due to the element specificity of resonant scattering, XRMR is sensitive only to magnetism in Pt.
Fig.~\ref{fig:ReflandSA}(d) shows the XRMR total specular reflectivity 
at $T=50$~K. No variation in reflectivity with $T$ was discernible. The XRMR spin-asymmetry averaged over measurements with positive and negative (saturating) $H$ (see~\onlinecite{suppl}) is shown at various $T$ in Fig.~\ref{fig:ReflandSA}(e). 
XRMR reflectivity and spin-asymmetry at each $T$ were fitted simultaneously using \texttt{GenX}, with sSLD consistent with XRR and PNR. The best-fits 
are shown as solid lines in Fig.~\ref{fig:ReflandSA}(d) and (e). 

The mSLD profiles that yield the best-fits to the XRMR data are shown in Fig.~\ref{fig:ReflandSA}(f). These mSLD profiles allow direct quantitative comparisons, but do not directly yield an absolute value for $M_\mathrm{Pt}$~\cite{suppl,Klewe2016}. At $T<T_\mathrm{C2}$ PIM is present in both Pt layers, and is larger in the upper Pt layer. PIM decays away from the interface with FM into the Pt layers over a lengthscale of $\sim 1$~nm at all $T$. PIM also reduces with increasing $T$. At $T=150$~K (above $T_\mathrm{C2}$) where the upper CFTB interface is paramagnetic, and at $T=250$~K (above $T_\mathrm{C1}$) where the entire
CFTB layer is paramagnetic, small mSLD contributions are found at the upper and lower interfaces, respectively. These indicate the experimental uncertainty in PIM amplitude derived from XRMR.

PIM, which is strongest at the upper interface, rapidly reduces with increasing $T$ and vanishes above $T_\mathrm{C2}$. Although smaller at low $T$, the lower Pt/CFTB interface PIM persists to significantly higher $T$ than that at the upper CFTB/Pt interface and vanishes at $T_\mathrm{C1}$. This suggests that PIM does scale with $M_\mathrm{CFTB}$ at a given interface, similar to Eq.~\ref{MvM}. The first important result presented here is that PIM requires the adjacent layer to have non-zero $M$, confirming that PIM in HMs does not arise solely due to interfacial charge-transfer~\cite{Ma’Mari2015}.

In the temperature region where both CFTB interfaces are ferromagnetic, the upper interface has a larger PIM but a smaller $M_\mathrm{CFTB}$, whereas the lower interface has a larger $M_\mathrm{CFTB}$ but a much smaller PIM. Such PIM asymmetry has been found in Pt/Co/Pt and Pd/Co/Pd, where $M_\mathrm{Co}$ should not differ significantly between the two interfaces. 
Such asymmetry, where different PIM is induced for similar $M_\mathrm{FM}$ at the HM/FM and FM/HM interfaces, is currently unexplained.



To quantitatively investigate the proposed scaling of PIM with $M_\mathrm{FM}$, in Fig.~\ref{fig:PIMvsM}\ we parametrically plot the amplitude of the XRMR mSLD (PIM) against the amplitude of the PNR mSLD ($M_\mathrm{FM}$) for each interface as a function of $T$. The conventional description of PIM, Eq.~\ref{MvM}, describes a proportionality between PIM and $M_\mathrm{FM}$.
For the lower Pt/CFTB interface the scaling is indeed linear but PIM vanishes whilst significant, non-zero, $M_\mathrm{CFTB}$ remains. For the upper CFTB/Pt interface there are only two measurements in the FM phase. Assuming linear scaling PIM again vanishes with similar, non-zero, $M_\mathrm{CFTB}$. For comparison, best fits to the XRMR data where PIM is constrained to scale with $M_\mathrm{CFTB}$, as described in Eq.~\ref{MvM}, are shown in the supplementary material, Ref.~\onlinecite{suppl}. This conventional model does not describe the data well. This direct measurement of the unexpected scaling of PIM with FM interface magnetization is the key experimental result presented in this manuscript.

Neglecting any $T$ dependence of the susceptibility $\chi$, which in Pt is less than 5~\%\ between 50~K and 200~K~\cite{Hoare1952}, we can modify Eq.~\ref{MvM} to describe the data in Fig.~\ref{fig:PIMvsM}\ as
\begin{equation}
\mathrm{PIM} = \chi \left( H_\mathrm{ex} - H_\mathrm{ex}^0\right) = \chi J \left( M_\mathrm{FM} - M_\mathrm{FM}^0\right),
\end{equation}
where the interface magnetization in the FM required to initiate PIM corresponds to a contribution $H_\mathrm{ex}^0 = J M_\mathrm{FM}^0$ ($J$ dimensionless) to the effective `exchange field' $H_\mathrm{ex}^\mathrm{eff} = H_\mathrm{ex} - H_\mathrm{ex}^0$. Here $H_\mathrm{ex}^0$ encapsulates the more complex behaviour of 
the spin-polarized 3$d$-5$d$ hybridization across the FM/HM interface. $M_\mathrm{FM}^0$ does not describe a magnetic dead-layer.

This new phenomenological scaling model shows that the effective susceptibility $\chi^\mathrm{eff} = \chi J$, which is proportional to the slope of the linear trendlines in Fig.~\ref{fig:PIMvsM}, is responsible for the PIM asymmetry. Indeed, $\chi^\mathrm{eff}$ is found to differ between upper and lower Pt layers. The final important result here is the demonstration that larger PIM at the upper interface results from the presence of different effective magnetic susceptibilities in the HM layers, and does not require asymmetry in the FM interface magnetization.

This asymmetry in $\chi^\mathrm{eff}$ may arise due to the local structure at the interfaces, via interfacial roughness and/or intermixing~\cite{Ferrer1997}. Intermixing should be greater at the lower HM/FM interface~\cite{Larde2009}, which would enhance $\chi^\mathrm{eff}$ in the lower HM layer, in contrast to the observed asymmetry. Our measurements reveal no structural mechanism for $J$ to vary, showing no significant difference between the interface widths for Pt/CFTB and CFTB/Pt interfaces. As the effective susceptibility may be modified by other means beyond interfacial intermixing, $J$ does not need to differ between these interfaces for different PIM to occur. Decreasing the size of Pt and Pd nanoparticles can significantly decrease $\chi$~\cite{Leeuwen1994}, which may also apply to polycrystal grain size. Lattice strain variation between the upper and lower HM layers may modify $g(E_\mathrm{F})$, and hence $\chi$~\cite{Hong2007}. These mechanisms may be plausible here, but cannot explain the asymmetry in strongly textured Ta/Pd/Co/Pd films with columnar grains as described in Ref.~\onlinecite{Kim2016}. Our measurements reveal how the PIM asymmetry arises, but a general mechanism explaining the difference in $\chi$ between upper and lower HM layers remains elusive.

\section{Conclusions}

To conclude, proximity-induced magnetism in a heavy metal and interface magnetization in a ferromagnet are shown to scale in a manner inconsistent with the expected phenomenological description; a small threshold magnetization in the ferromagnet is required to begin to then induce magnetism in the adjacent heavy metal. In a trilayer, the proximity-induced magnetism is shown to be larger in the upper heavy metal layer as a result of the heavy metal layers having different effective magnetic susceptibilities. This demonstrates how asymmetric PIM arises in HM/FM/HM trilayers.





\section*{Acknowledgements}

We acknowledge support from Nigerian government TETFund scheme (OI), EPSRC DTP (BN and RMR-R), Republic of Turkey Ministry of National Education (MT), EPSRC Grant Ref.\ EP/L000121/1, HEFCE ODA,  and the Royal Society (ATH). XMaS is a UK national facility supported by EPSRC. We thank STFC for ISIS beamtime RB1620118 and access to ISIS R53 Materials Characterization Lab, and D.\ Atkinson, D.M.\ Burn, C.\ Cox, T.P.A.\ Hase, P.D.\ Hatton, and K.\ Morrison for stimulating discussions. Supporting data are available at http://dx.doi.org/10.5286/ISIS.E.86389562 and http://dx.doi.org/10.15128/r10v8380575.

\bibliographystyle{apsrev4-1}

%

\end{document}